\documentclass{ws-procs9x6}

\begin{document}

\title{Measurement of Prompt Photon in $\sqrt{\lowercase{s}}=200$G\lowercase{e}V \lowercase{pp} collisions}

\author{K. Okada for the PHENIX collaboration}

\address{RIKEN-BNL Research Center,\\
Brookhaven National Laboratory, \\
Upton, NY 11973-5000, USA \\
E-mail: okada@bnl.gov}

\maketitle

\abstracts{
This report presents the preliminary result for 
the prompt photon production cross section in proton-proton 
collisions in the mid-rapidity region using the PHENIX detector.
The NLO pQCD calculation is in good agreement with the data.
The measurement was made with and without an isolation cut.
The isolation cut significantly 
improves the signal purity without reducing the signal yield.
This is an important 
step for the future spin asymmetry measurement.
}

\section{Introduction}
Prompt photon production in proton-proton collisions is a good probe of the parton structure in the proton. Its leading sub-process is gluon-quark Compton scattering. 
In polarized proton-proton collisions at RHIC, it is a good tool to access the spin structure function of the gluon in the proton. 
Two types of processes contribute to the prompt photon production cross section:
the so-called 'direct' component, where the photon is emitted via a point like (direct) 
coupling to a quark, and the fragmentation component, in which the photon originates 
from the fragmentation of a final state parton.
In the latter case, hadronic activities are expected 
to be accompanied with the photon. 
These processes are thought to be separated by applying an isolation cut on the photon.
The cross section of the prompt photon at the mid-rapidity region in proton-proton collisions at $\sqrt{s}=200$GeV  has been measured \cite{run2photon} using the PHENIX detector.
In this paper, we report the measurement with increased statistics, and 
an isolation cut on photons is applied for the first time.

\section{Experimental Setup}
\begin{figure}[ht]
\centerline{\epsfxsize=2.0in\epsfbox{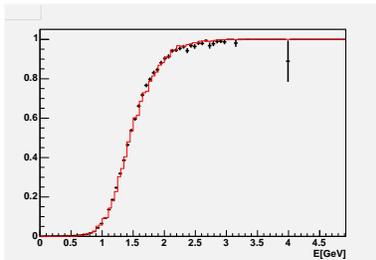}}   
\caption{Turn-on behavior of the EMCal trigger. The line shows the expected behavior.}
\label{fig:turn-on}
\end{figure}

The analysis is based on the integrated luminosity of 266 $\rm{nb}^{-1}$ collected by the PHENIX detector \cite{PHENIXNIM} at the RHIC proton-proton run in May 2003.
Photons are detected by electromagnetic calorimeters (EMCal) 
in the central arms, each of which has an azimuthal coverage of $90^\circ$ 
and a pseudo-rapidity ($\eta$) coverage of $\pm0.35$.
In this paper, we report only on the measurement done with one arm.
The EMCal has such a very fine granularity (10$\times 10 \rm{mrad^2}$) that 
we can efficiently identify background photons by reconstructing $\pi^0$'s. 
The energy scale and resolution were tuned by checking the $\pi^0$ mass peak position and its width.
In addition PHENIX has a tracking system in front of the EMCal which was useful to reject background clusters produced by charged hadrons.
Data were collected with a minimum bias trigger provided by the beam-beam counters (BBC) placed in $3.0\!< \!|\eta|\!<\!3.9$ region and with an EMCal trigger.
By requiring a high energy cluster with the EMCal trigger, event condensation by a factor of 120 was achieved. 
Fig. \ref{fig:turn-on} shows that the trigger efficiency attains a plateau
at 100\% for single photons above $E\sim3\rm{GeV}$ for the active channels of EMCal.

\section{Analysis Procedure}
 The prompt photon signal is obtained by subtracting known backgrounds from all EMCal clusters.
 The contribution of hadronic interactions is rejected by a photon shower shape cut and by requiring that no charged track is associated with the cluster. 
 The dominant source of the background is two photon decays of $\pi^0$'s. 
This background is estimated by reconstructing $\pi^0$'s from two photons.
A photons is rejected as a $\pi^0$ decay photon when the invariant mass of the photon and another photon in the same event is consistent with mass of $\pi^0$. 
The accidental coincidence is taken into account using the vicinity of $\pi^0$ mass window. 
If only one of the two decay photons from a $\pi^0$ is detected, it mimics a prompt photon signal. 
This probability depends mainly on geometrical acceptance and kinematics of $\pi^0$ decays, which can be well reproduced by a Monte-Carlo simulation.
Thanks to the highly segmentated EMCal, two photon clusters from a $\pi^0$ are well separated in the $p_T$ region of this measurement.

Besides the simple subtraction method, an isolation cut is applied to prompt photon candidates. 
It requires that the energy sum ($E_{sum}$) in an angular cone around a photon is less than a certain fraction of the photon energy (Eq. \ref{eq:isocut}).
\begin{equation}
E_{sum}(R\!<\!0.5)\!<\!E_\gamma \!\times\! 0.1,\:\: R=\!\sqrt{\Delta \eta^2+\Delta \phi^2} 
\label{eq:isocut}
\end{equation}

The amount of photons from hadronic decays other than $\pi^0$'s is estimated based on the total amount of $\pi^0$'s. 

In the subtraction method,
of all photon clusters, 50\% (20\%) are tagged as photons from $\pi^0$, and 85\% (30\%) are estimated as photons 
from all hadrons including missing $\pi^0$'s at $p_T=5\,\rm{GeV}\!/\!\it{c}$ $(16\,\rm{GeV}\!/\!\it{c})$.
In the isolation method, ratios are improved to 25\% ($<\!5\%$) and 65\% ($<\!5\%$). 
The uncertainties of the energy scale, of the $\pi^0$ extraction, and of 
the hadron-to-$\pi^0$ production ratio are sources of systematic error.

After the yield extraction, correction factors such as acceptance, efficiency, luminosity, and BBC trigger bias are applied to calculate the cross section. 
\section{Results and discussion}

\begin{figure}[ht]
 \begin{minipage}[h]{0.45\linewidth}
     \mbox{\epsfxsize=2.0in\epsfbox{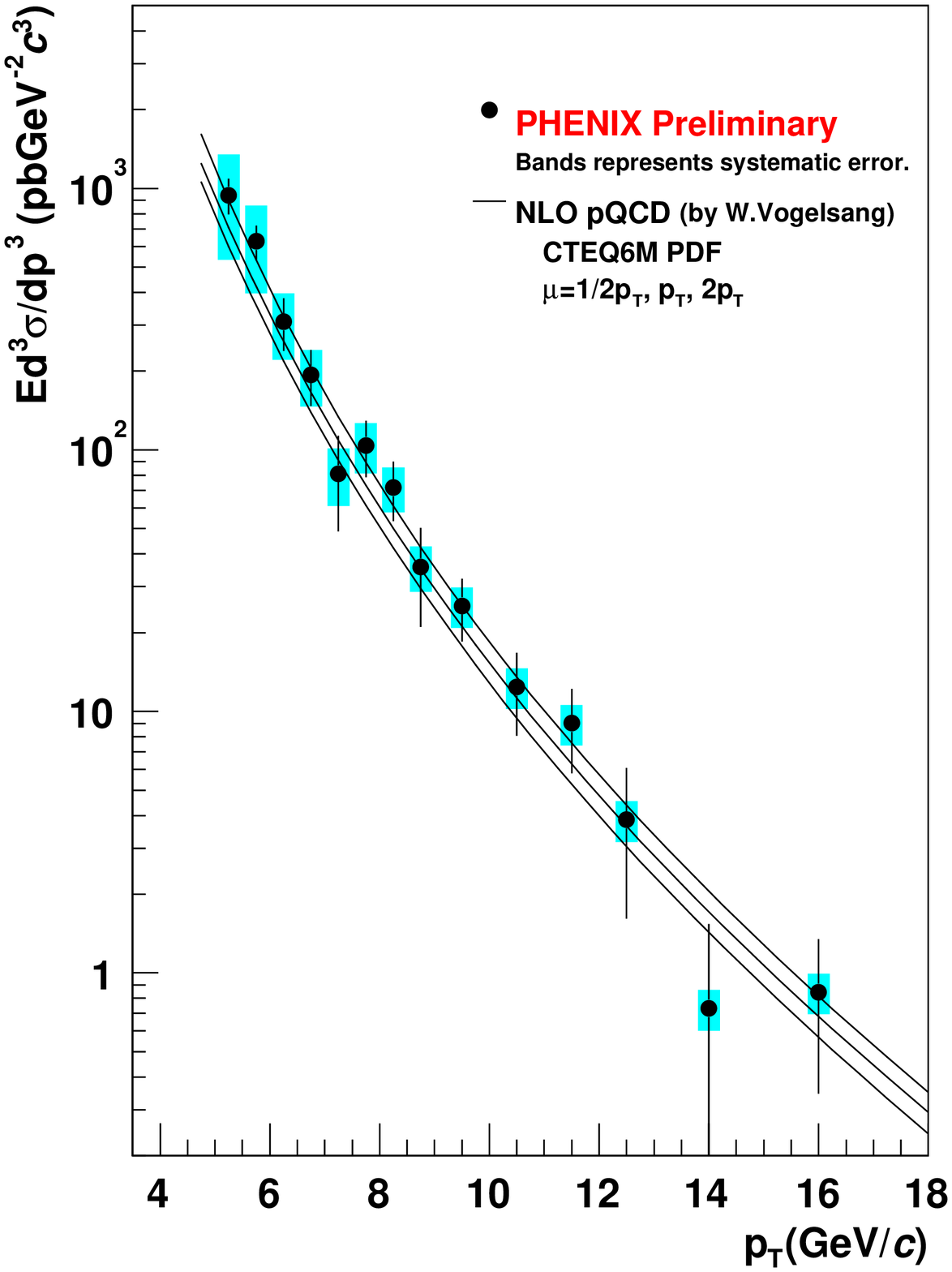}}   
 \end{minipage}
 \hfill
 \begin{minipage}[h]{0.45\linewidth}
     \mbox{\epsfxsize=2.0in\epsfbox{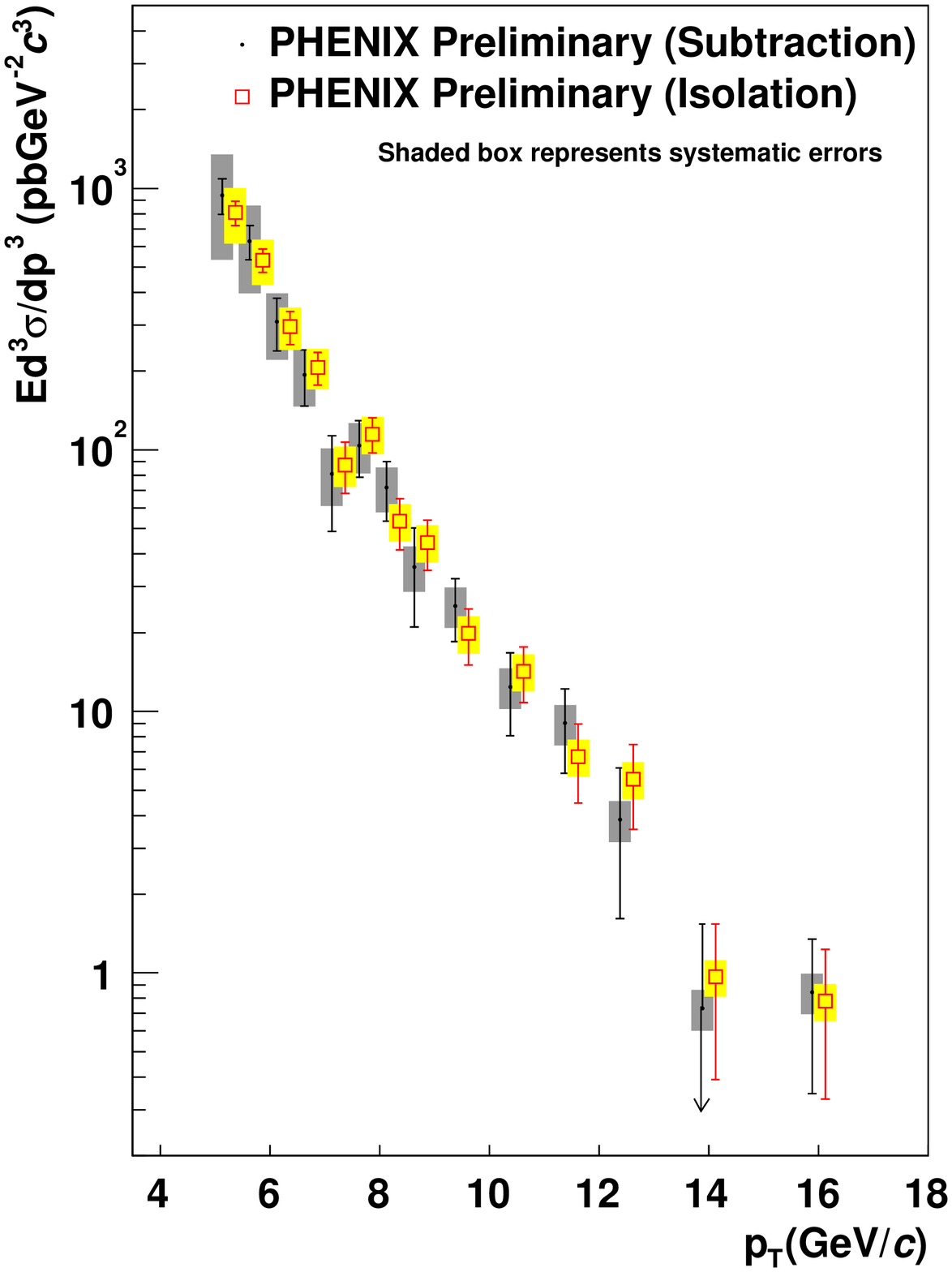}}   
 \end{minipage}
\caption{[left] Result of the subtraction method with NLO pQCD calculation. 
[right] Comparison to the isolation method.}
\label{fig:xsec}
\end{figure}

Fig. \ref{fig:xsec} (left) shows our measurement of the prompt photon cross section as a function of $p_T$. Bands correspond to systematic errors which are listed in 
Table \ref{table:syserr} for our lowest and highest $p_T$ bins.
Curves show a NLO pQCD calculation with 3 different scales \cite{werner}. 
The theory agrees well with our experimental measurement.
In Fig. \ref{fig:xsec} (right), the two analysis methods are compared and no significant reduction in
the isolation method is observed. 
It suggests either (1) the 'direct' component is the dominant source of the 
prompt photon or (2) the reduction of the 'fragment' component by the 
isolation cut is small.

In summary, we have measured the cross section of prompt photon in proton-proton collisions at $\sqrt{s}=200$ GeV. 
The NLO pQCD calculation agrees with the measurement very well. 
The isolation method significantly enhances the signal purity.
These results are important for future spin asymmetry measurements of this channel.

\begin{table}[ht]
\tbl{Systematic error table}
{
\footnotesize
\begin{tabular}{crr|rr}
\hline
{} &{} &{} &{} &{}\\[-1.5ex]
{} & \multicolumn{2}{c|}{Subtraction} & \multicolumn{2}{c}{Isolation} \\
{} & Lowest & Highest & Lowest & Highest\\
{} & $5-5.5$ & $15-17$ & $5-5.5$ & $15-17$\\
{} & [GeV/$c$] & [GeV/$c$] & [GeV/$c$] & [GeV/$c$] \\
\hline
{} &{} &{} &{} &{}\\[-1.5ex]
$\pi^0$ photon estimation      &30\% &5  &16 &2 \\
Non $\pi^0$ photon estimation  &27   &6  &8  &1 \\
Photon acceptance and smearing &10   &10 &10 &10\\
Photon conversion effect       &1    &1  &1  &1\\
Luminosity measurement         &12   &12 &12 &12 \\
BBC trigger bias               &3    &3  &3  &3\\
\hline
{} &{} &{} &{} &{}\\[-1.5ex]
Total                          &43\% &18 &24 &16\\
\hline
\end{tabular}
\label{table:syserr}
}
\vspace*{-13pt}
\end{table}

\end{document}